\documentclass[aps,prb,reprint,showpacs,superscriptaddress,twocolumn]{revtex4-1}
\usepackage{amsfonts}
\usepackage{graphicx}
\usepackage{amsmath}
\usepackage{amssymb}
\usepackage[breaklinks=true,dvips]{hyperref}
\usepackage{breakurl}
\usepackage[utf8x]{inputenc}
\usepackage{mathrsfs}
\usepackage{color}

\hypersetup{
  colorlinks=true,linkcolor=blue,citecolor=blue,
  filecolor=blue,urlcolor=blue,breaklinks=true
}
\begin{document}

\title{Quantum teleportation between two quantum spin Hall insulator rings}
\author{Thiago Prud\^encio}
\email{prudencio.thiago@ufma.br, prudncio@chalmers.se}
\affiliation{Coordination of Science and Technology - CCCT/BICT, Federal University of Maranh\~ao - UFMA,
%Campus Bacanga, 
65080-805, S\~ao Lu\'is-MA, Brazil.}
\affiliation{Department of Physics, University of Gothenburg/Chalmers University of Technology
, SE 412 96 Gothenburg, Sweden.}

%\date{\textcolor{blue}{\it \today %november 2019 %June 21 2015
%}}

\begin{abstract}
A quantum teleportation between two quantum spin Hall insulator (QSHI) rings is proposed. A
qubit state is generated in the first QSHI ring and
the second is used as a quantum channel in entangled QSHI state. In the intersection of outer edges is placed a joint detector, allowing a Bell state analysis. After the Bell state measurement, Alice, that is composed originally of a qubit in the first QSHI ring and one spin state in the second QSHI ring, allows Bob,
that is composed of the other spin edge state in the second QSHI ring, to realize the corresponding
control of the tunneling amplitudes in the second QSHI ring, what implements a complete quantum teleportation between the QSHI rings. It is also showed that Bob's table,
used after the classical communication by Alice, is the same independently
the way the qubit is generated in the first QSHI ring. This description indicates that a quantum
teleportation can be fully implemented in the context of two-dimensional (2D) topological edge
states and opens perspectives in the scenario of quantum information in topological matter,
decoherence free systems, as well as for the development of quantum technologies associated to
spintronics and quantum computation.
\end{abstract}
%%. %with robustness against decoherence. 
%As a quantum device, the two QSHI ring device Due the elements that compose this protocol are realistic and achievable in solid-state devices, it can be realized experimentally and can also be used as an element for
%\pacs{03.67.Ac,03.67.Hk,03.67.Lx}
\maketitle

\section{Introduction}
%\begin{widetext} 
%\end{widetext}
%I read the Introduction & Concluding remarks of the paper. Look fine overall, but I would suggest you to (a) remove the reference to Star Trek (!) (doesn't quite fit in this type of paper); (b) move the more technical parts of the teleportation protocol (where you use formulas) to  Sec VI, (c) add some more references for background and related papers*; (d) run a spell check; (e) remove "masterpiece" in "masterpiece protocol" (sounds funny in my ears).

%*For experimental facts about QSH device performance, you could use

Quantum spin Hall insulator (QSHI) is a %distinctive 
topologically non-trivial state of matter, allowing %a
dissipationless spin current \cite{mukarami2004}, characterized by an insulating two-dimensional (2D) bulk with conducting helical Luttinger liquids at its boundary, forming Kramers' partners, where opposite spin modes counterpropagate \cite{kane2005,bernevig2006p}. %and described by a Dirac massless fermion theory in $1+1$ dimension. 
Due the strong spin polarization of the gapless edge carriers, provided by the helical spin-orbit (SO) coupling that attaches spin and momentum \cite{yang2006}, non-necessity of an external magnetic field \cite{murakami2006}, and robustness against decoherence, characterized by the absence of backscattering interactions \cite{xu2006}, the realization of the QSHI phase is associated to a progress in the field of spintronics \cite{lombardi2009}. It can be classified by means of a topological $Z_{2}$ index \cite{kane2006}. QSHI state has been showed to be realizable in HgTe/(Hg,Cd)Te quantum wells \cite{bernevig2006}, with experiments in optical and electron-beam lithography used to fabricate inverted structure quantum well in regular Hall bar shape  geometry in different sizes where the condutance showed the edge transport expected to a QSHI above and at the critical thickness of $6.3$nm, with a plateau of residual condutance close to $7.75 \times 10^{-5}\Omega^{-1} \approx 2e^{2}/h$, at temperature below $10$K \cite{konig2007}. %This experiment showed that such an state can be
%
%Since this effect occurs in the absence of magnetic field, it can have an important role in the field of spintronics, allowing direct manipulation of spin. 
%The spin current can flow without dissipation. 
%Allow the fabrication of spintronic devices with low power dissipation. Kramers' doublets.
%
%It is a topological non-trivial state of matter. QSHI is invariant under time reversal, with charge excitation gap in the bulk and gapless edge states topologically protected in the edge. Two states with opposed spin polarization counterpropagate in a given edge. Spin polarized Luttinger liquids with momentum locked to spin
%
The investigation of tunneling transport of these helical states between different edges, by a point contact controlled by a gate voltage \cite{strom2009} or at corner junctions \cite{hou2009} revealed that the electron-electron interactions do not interfere in the helicity characteristic of the QSHI phase, allowing the construction of more elaborate devices involving QSHIs. 
%
%In the quantum spin Hall effect conducting polarized spin channels are developed where carries with opposed spin move in opposed directions, locking the spin direction to the momentum direction in the edges of the sample. The presence of tunneling junction then does not affect such a scenario.

%In the quantum Hall effect conducting channels develop in the edges of the sample in the presence of a magnetic field, making the transverse electrical condutance of the material quantized. These charge carries resistent to scattering and can be transported without dissipation in the channels. For this reason, they are very promissing in integrated circuit technology and spintronics.

%The quantum spin Hall (QSH) effect \cite{bernevig2006} is particular example of a two-dimensional (2D) topological insulator, particularly interesting due the existence of Luttinger liquids in the edges and a strong spin-orbit
%coupling that makes the conducting states helical in the edge [ref].

%Quantum spin Hall insulator (QSHI) 
Particularly, QSHI ring is a quantum device that implements appropriately a QSHI state and can be used for integration in quantum devices, quantum technologies and nanostructures \cite{whitesides2005,cao2019}. Endowed with inner and outer edges where spin edge states can flow in the edges as Luttinger liquids, or more properly, helical liquids \cite{wu2006}, while the bulk is kept in a highly insulating state, i.e., with a gap that cannot be easily closed, the presence of tunneling junctions between the inner and outer edges allows for the interchange between inner and outer edge states while in the other regions, the critical thickness separates the two conducting edges by means of the insulating bulk \cite{hou2009}. 

The injection of spin particles by means of a source in one of the edges imply in a spin current whose flow is resistent against backscattering interaction, meaning that a given spin state cannot be inverted without to be transformed in its Krammers' partner. They are protected under time-reversal (TR) symmetry. This property is particularly important in quantum information and quantum computation contexts, since it makes the state robust against decoherence \cite{zurek2003}, making the edge modes are protected \cite{groh2016}. The effect is exactly to forbid a backscattering of the spin state, keeping its coherence \cite{buttiker1988}. This feature is a particular advantage when dealing with quantum information protocols, particularly in the case of quantum teleportation.
%paragraphy 4 - reply

The properties of QSHI ring are then a consequence of its strong spin-orbit coupling of the edge states in a QSHI state. This helical property, means that clockwise and counterclockwise propagations always occur with opposite spins \cite{strom2010}. Other important aspect of QSHI rings is the compact form, important in technological devices as nano-size components, including bioinspired computing \cite{grollier2016}. The experimental size of a QSHI ring device can be estimated from the litographic techniques used for such materials \cite{konig2007}. Internal radius could be of order of $130$ nm % nm127,32 nm$ 
while the external radius of size $230$nm. Tunneling junctions can be of order of $20$ nm and detectors can be disposed in the interfaces with size of order from $20$nm to $50$ nm. This device then has a estimated total area %of $230 \time 450$ nm$^{2}$, what 
that is close to other devices fabricated in nanosize order, then being technologically achievable \cite{kawakami2015,lin2016}. 
These dimensions can also be potentially improved with science and technology progress in this field.
%and consequently QSHI rings are a very promissing as a quantum device to implement both quantum information [ref] and spintronics [ref]. Other important applications of QSHI rings are in the study periodic quantum dynamics [ref] and geometrical phases [ref] in the context of QSH effect.

Bipartite entanglement state in a QSHI ring has been proposed with respect to generation and control, taking into account the electric control of tunneling junctions between the inner and outer edges \cite{strom2015}. The two QSHI ring setup discussed in this paper is a result of the single QSHI ring discussed in \cite{strom2015}. By allowing the generation of maximally entangled states and the realization a Bell tests in the scenario of QSHI rings, a vast possibility of applications can be glimpsed in the context of quantum information \cite{nilsen,streltsov}. 

The first application is the use of the QSHI ring as a quantum channel, this allows the QSHI ring be used as a quantum device that makes the way to quantum communication in spintronics. A second application, that is the fundamental in the storage part of this device, is the generation of a qubit state with the edge states. This requires a quantum colapse of one mode, i.e., single spin detection of one of the edge states in order to implement a qubit of edge states. %Both of these applications will be discussed in this paper. %the possibility of using this device as a quantum information device [ref], particularly in the aspect of generation of maximally entangled states in 2D topological insulators [ref]. In the case of quantum teleportation, this feature can be considered a good resource for realizing a quantum channel. One other aspect that was not highlited in [ref] was that the QSHI ring can also be used to generate a qubit state, by means of a spin detection of one mode of the two-spin edge states. By collapsing one of two-particle states properly by the introduction of a projection detector, the whole QSHI ring is put in a qubit state of the remaining edge state, that still stays robust against decoherence. %This is a particular since it is still a the presence of a postselector detector only changes a phase in the total state. 

As a consequence, a QSHI ring is a proper device to build elements of quantum information and quantum computation. A third application requires the integration with other devices or other QSHI rings. This is necessary to establish the integration with other elements of quantum information. In particular, in the case of a device made of two QSHI rings it is possible to deal with first and second possible application mentioned above: the presence of both quantum channel and qubit. 

This is the precise case discussed in this paper, where a two QSHI rings device will be used to establish a quantum teleportation between the QSHI rings. In the case of quantum teleportation, the building elements are essentially the presence of an entangled state and a qubit state. As such, a quantum device for realizing a quantum teleportation will require precisely two QSHI rings. This setup has advantages with respect to optical and other material devices when the process occur in nanoscale, since the robustness against coherence favors the QSHI device.

\begin{figure}[h]
\centering
\includegraphics[scale=0.35]{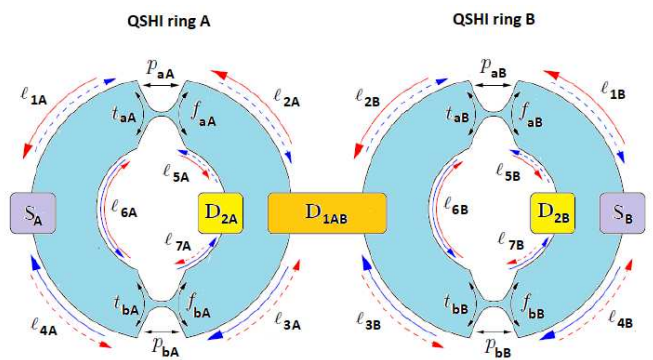}
\caption{(Color online) Setup of two QSHI rings $A$ and $B$, composed of the sources $S_{A}$ and $S_{B}$, detectors $D_{2A}$and $D_{2B}$, joint detector $D_{1AB}$ and tunneling junctions $aA$, $bA$, $aB$ and $bB$, in which are associated tunneling amplitudes $t_{aA}$, $t_{bA}$, $t_{aB}$, $t_{bB}$ (with spin preserved in the opposed edge), preserving amplitudes $p_{aA}$, $p_{bA}$, $p_{aB}$, $p_{bB}$ (at the same edge) and flipping amplitudes $f_{aA}$, $f_{bA}$, $f_{aB}$, $f_{bB}$. Spin up is represented by red line, spin down is represented by blue line. In the figure, the arrow represent the clockwise direction and counterclockwise direction associated to the respective spin due strong spin-orbit coupling.}
\label{2rings}
\end{figure}

The configuration of the scenario of two QSHI rings device is displayed in figure \ref{2rings}, that will be detailed in the next sections. In the case of quantum teleportation in such device a new challenge comes into play and will be discussed in detail in this paper. The quantum teleportation requires a quantum correlation between the two QSHI rings. Such a correlation is generated by means what is called a Bell state analysis \cite{weinfurter1994,braunstein1995,michler1996}. This was the fundamental step to implement experimentally the quantum teleportation for the first time \cite{bouwmeester1997}. As such, from the practical point of view, in order to implement a quantum teleportation in the two QSHI ring device, the quantum communication between the two QSHI rings will require a joint detector that establishes Bell state analysis.
%
%In order to allow quantum teleportation 

In the present proposal, a two QSHI rings device is proposed, in which two QSHI rings $A$ and $B$ can be correlated by the presence of a joint detector in their outer edges, as displayed in the figure \ref{2rings} in order to implement quantum teleportation in this nanoscale device. The joint detector will implement a Bell state analysis that will be discussed in the Sec VIII. In section II, a review of the general protocol of quantum teleportation is discussed. The details of the two QSHI rings device and the figure the figure \ref{2rings} will be discussed in detail in the Sec III. The hamiltonian of the system is discussed in Sec IV and the scattering matrices are derived in Sec V. Quantum states in QSHI rings are discussed in Sec VI. A quantum teleportation between the two QSHI rings will be implemented in this system, with the detailed protocol to be discussed in Sec VII. Finally, Sec XI is reserved to concluding remarks.

\section{General Description of Quantum teleportation}

Quantum teleportation is one of the most promissing and fundamental quantum information protocols, with several implementations for quantum technologies \cite{pindarola2015,vedral2007}. This protocol allows the realization of communication using non-locality, allowing its implementantion in cases in which Alice and Bob, in quantum information jargon, are really far from each other \cite{ren2017,pfaff2014}. Just to review how this protocol works, quantum teleportation was defined originally \cite{bennett1993} as a quantum communication between Alice and Bob, in which Alice has a qubit state and shares with Bob a quantum channel, that in this case is a bipartite quantum entangled state. The name teleportation comes from the fact that Alice's qubit is destroyed in the final step of the protocol and is exactly reconstructed in Bob's side, %as a Bob's qubit
 in the exactly same way as Alice's qubit%, similar to what occurs in the teleportation of Star Trek [ref], from where this name commes, reason as it is called quantum teleportation
. This feature in quantum teleportation is not only a choice of the protocol, but a consequence of the no-cloning theorem \cite{wooters1982}.%, that forbids a perfect copy to Bob except in the case in which the original is destroyed [ref], making an analogy with computer science, it could be said that copy and paste is forbidden in quantum domain due to no-cloning theorem, while cut and paste is allowed. 

In the quantum teleportation protocol, Alice should make a Bell state analysis %joint measurement of Bell type [ref] 
in her states, resulting in an outcome of one of the four Bell states, $|\mathcal{B}_{1}\rangle,|\mathcal{B}_{2}\rangle,|\mathcal{B}_{3}\rangle,|\mathcal{B}_{4}\rangle$ \cite{weinfurter1994}, and since the total state is a superposition of states of type $|\mathcal{B}_{i}\rangle |\mathcal{R}_{i}\mathcal{Q}\rangle$, $1,2,3,4$, where $\mathcal{R}_{i}$ is one of four possible rotation matrices acting on the qubit state,  Alice's measurement project Bob's state in one of the four states $|\mathcal{R}_{i}\mathcal{Q}\rangle$, corresponding to a rotated state with respect to the original qubit. 

This result then requires the corresponding inversion of the rotation matrix $\mathcal{R}_{i}$. The presence of a classical channel is then required, since Alice should communicate to Bob what is her outcome in order to Bob implement the appropriate inversion by implementing the inverse of the operation by the rotation matrix $\mathcal{R}_{i}^{-1}$. This allows %to Bob implement the inversion  that allows 
to complete the quantum teleportation of the qubit $|\mathcal{Q}\rangle$. As a consequence, the quantum teleportation protocol requires the presence of a quantum channel and a classical channel in order to be successfully realized %completelly implemented 
with highest possible fidelity. Particularly, the classical channel is fundamental to avoid contradictions with the relativistic causality \cite{bohr1935}.
%
%Quantum teleportation protocols have been proposed, since the original scheme, in several contexts, for instance, in scenarios of quantum optics [ref], trapped ions [ref], solid state devices [ref], circuit QED [ref], and successfully experimentally realized in several contexts [ref]. Particularly, the experiments involving far distant quantum teleportation have been achieved by means of some breakthoughs as a quantum teleportation when Alice and Bob are separated by 143 km [ref], or when Alice is in earth while Bob is in a satellite around the earth [ref]. 
%These possibilities also openned a new scenario for realization of quantum networks [ref] and quantum internet [ref]. 
%On the other hand, 

In the context of quantum circuits and nano-devices, the quantum teleportation can be implemented in short and large distances with more efficiency than making the physical transport of the particles. %, that why quantum teleportation can also be considered a quantum computer primitive [ref].
The absence of dissipation in QSHI rings, makes this device ideal to allow communication between the two QSHI rings using %by means of 
quantum teleportation. This system also can have advantage with respect to optical devices \cite{barasinski2019}, by the fact it can be forwardly integrated in solid state technologies.%
%In the context of QSHIs 2D topological insulators and QSHI ring devices, there is a lack of theoretical and experimental proposals in the context of quantum information and quantum computation, pa

Since the implementation of a quantum teleportation requires at least a tripartite state, i. e., a quantum channel and  a qubit, a single QSHI ring is not appropriate to implement this protocol. 

On the other hand, the building of a two QSHI rings connected by a joint detector can be used as an ideal quantum device for realization of quantum teleportaion using edge states. %The joint detector in the outer edges

\section{Setup}

The setup starts with two QSHI rings $A$ and $B$. The bulk of each QSHI ring is in insulating state and is represented in blue in the figure \ref{2rings}. The edge states flow in the border between the blue and white region, i.e., the frontier between the QSHI rings and their outside, forming a two-dimensional (2D) topological insulator exhibiting a QSHI effect. The QSHI rings $A$ and $B$ are endowed with inner and outer edges that are connected by two tunneling junctions in each one. The sources $S_{A}$ and $S_{B}$ inject particles in each QSHI rings $A$ and $B$, respectively. Detectors in the inner edges are represented by $D_{2A}$ and $D_{2B}$ for the QSHI rings $A$ and $B$, respectively. In the outer edges a detector $D_{1AB}$ is common to both QSHI rings and it will be responsible for realizing joint operations that can affect simultaneously both QSHI rings $A$ and $B$. %While $D_{2A}$ and $D_{2B}$ are responsible to make postselection and projection in a single state, the $D_{1AB}$ should realize a single state postselection and a two particle joint measurement.

Geometrically, the QSHI rings $A$ and $B$ have edges of a given length. In the figure \ref{2rings}, the lengths  $l_{1A},l_{2A},l_{3A},l_{4A}$ are outer edge lengths  of the QSHI ring $A$, while $l_{5A},l_{6A}$ and $l_{7A}$ are are the inner edge lengths  at the same QSHI ring. Similarly, for the QSHI ring $B$, the outer lengths are represented by $l_{1B},l_{2B},l_{3B},l_{4B}$ and the inner edge lengths are represented by $l_{5B},l_{6B}$ and $l_{7B}$. Although the absence of backscattering forbids the edge states of reverse spin staying in the same edge, these states can acquire geometrical phases proportional to the corresponding outer or inner edge lengths and to the respective particle momenta $\pm K_{A}$, $\pm K_{B}$, to the QSHI rings $A$ and $B$ %, dependent on the energy of each  Krammer's partners $E_{A}$ and $E_{B}$ %corresponds to the energy of the in the ring $B$
, respectively. 

The manufacturing of the two QSHI rings device should be nano-size, capable of being part of a quantum computer or quantum information device, used in the present case for the implementation of quantum teleportation. The material should be an 2D topological insulator at QSHI state, consequently it should be used in the corresponding temperature in which the material achieves such this topological phase. The experimental size of these two QSHI rings device can be estimated from the litographic techniques, as estimated in the case of a single QSHI ring in the previous section. The internal radius could be of order $R_{1A}=R_{1B}= 130$ nm % nm127,32 nm$ 
while the external radius of size $R_{2A}=R_{2B}=230$nm. Tunneling junctions can be of order $t_{J}=20$ nm and the detectors in the interfaces $D_{2A}=D_{2B}=20$nm and $D_{1AB}=50$ nm. This device then has a estimated total area of $230 \time 450$ nm$^{2}$. This could be a very important advance in the context of quantum devices and can be improved with advances in technology and experimental techniques.

The interchange between the positions from inner or outer edges are only allowed by the presence of the tunneling junctions in each QSHI ring. The tunneling junctions are represented by $aA$, $bA$, $aB$, $bB$, where the small letters $a$ and $b$ corresponds to the two tunneling junctions of each QSHI ring $A$ and $B$, respectively. Each QSHI ring is identified by the capital letters of the corresponding QSHI rings $A$ and $B$, displayed in the figure \ref{2rings}. The processes in the tunneling junctions are only of three types, associated to the corresponding probability amplitudes. 

The probability of preserving the state in the same edge after scattering the corresponding tunneling junction for the QSHI rings $A$ and $B$ are represented by $p_{aA}$, $p_{bA}$, $p_{aB}$ and $p_{bB}$, while the probability of tunneling in the junction to the other edge are given by $t_{aA}$, $t_{bA}$, $t_{aB}$ and $t_{bB}$, when the spin is not flipped in the process and $f_{aA}$, $f_{bA}$, $f_{aB}$ and $f_{bB}$, when the spin is flipped in the process. %also represented in the figure \ref{2rings}. 
Due the conservation of total probability, these amplitudes should satisfy 
\begin{eqnarray}
&& t_{aA}t_{aA}^{*} + p_{aA}p_{aA}^{*} + f_{aA}f_{aA}^{*} =1, \\
&& t_{bA}t_{bA}^{*} + p_{bA}p_{bA}^{*} + f_{bA}f_{bA}^{*} =1, \\
&& t_{aB}t_{aB}^{*} + p_{aB}p_{aB}^{*} + f_{aB}f_{aB}^{*} =1, \\ 
&& t_{bB}t_{bB}^{*} + p_{bB}p_{bB}^{*} + f_{bB}f_{bB}^{*} =1, %\\ 
\end{eqnarray}
where the $^{*}$ corresponds to the complex conjugation. These relations can be rewritten in terms of the equivalent form %squares  
\begin{eqnarray}
&& |t_{aA}|^{2} + |p_{aA}|^{2} + |f_{aA}|^{2} =1, \\
&& |t_{bA}|^{2} + |p_{bA}|^{2} + |f_{bA}|^{2} =1, \\
&& |t_{aB}|^{2} + |p_{aB}|^{2} + |f_{aA}|^{2} =1, \\ 
&& |t_{bB}|^{2} + |p_{bB}|^{2} + |f_{bB}|^{2} =1. %\\ 
\end{eqnarray}
Due the strong spin-orbit coupling produced in the QSHI state, the backscattering process is forbidden, as discussed in the first section, implying that the direction of propagation is attached to the spin direction, spin up for clockwise %direction 
and spin down for counterclockwise direction, and vice-versa. Although the spin-orbit producing helical states can be found in different materials, in the QSHI ring the strong spin-orbit coupling it is a consequence of the QSHI effect, where the only form of flipping the spin state is by means of a non-zero flipping amplitude of tunneling junction. This fact imply that the absence of the tunneling junctions makes impossible the generation of spin superpositions, qubits, in this device. 

Taking into account these points, the QSHI ring $A$ will be used to generate a qubit state that will be Alice's particle to be teleported to Bob, in quantum information jargon. The qubit will be generated after operating with the detector $D_{2A}$ in one of the states, labeled by $2$, generated by the source $S_{A}$, that will produce a two mode spin state corresponding to a Krammers pair $|\uparrow\rangle_{1,A}|\downarrow\rangle_{2,A}$. On the other hand, in the QSHI ring $B$, the source $S_{B}$ will inject at the same time as the source in the QSHI ring $B$ a Krammers pair $|\uparrow\rangle_{1,B}|\downarrow\rangle_{2,B}$ and will produce instead a quantum channel that will be used to implement the quantum teleportation of the qubit state to one of the modes in the QSHI ring $B$. The injected state in the QSHI ring $B$ will be then used after the generation of an entangled state. 

As a quantum channel, the QSHI ring $B$ will establish a communication channel between Alice and Bob, with the mode $1,B$ belonging to Alice and with the mode $2,B$ belonging to Bob. In this sense, the qubit in the QSHI ring $A$ and the first mode in the QSHI ring $B$ belong to Alice, while Bob has the second mode in the QSHI ring $B$. In this scenario, a quantum teleportation will be established between Alice and Bob, and consequently from the QSHI ring $A$ to the QSHI ring $B$, implementing the quantum teleportation, a quantum information protocol, in the context of QSHI rings.

In the case of this particular protocol, the classical communication between Alice and Bob is represented by the fact that after the joint operation in the QSHI rings $A$ and $B$, the set of tunneling junctions in the QSHI ring $B$ should be adjusted properly in order to generate the corresponding rotation in the superposition state generated Bob's side. As a consequence, a faithful quantum teleportation can be idealy implemented in the scenario of two QSHI ring devices% and, as we will discuss, 
using the possibility of experimental electric control of the entanglement \cite{strom2015}.

This proposed setup has some important points that makes the system in figure \ref{2rings} interesting for quantum information implementations. Since it is strongly robust against backscattering processes, due the own nature of the QSHI state, the strong spin-orbit coupling of the edge state, the superposition state generated is robust against decoherence effects, %are very weak, 
making the time in which the superposition state is maintained much larger than the decoherence time. On the other hand, since the generation of QSHI state is a very challenging, requiring a state-of-the-art technology to be implemented. These devices should be part of a progress in the technology associated to topological insulators and QSHI systems. Once these devices can be built and quantum information protocols can be implemented in it, promising applications integrates with systems in solid state systems, as circuit QED implementations, can be imagined, with the particular advantage that QSHI device is practically decoherence free. 

\section{Hamiltonian}

%\subsection{Setup}

%\subsection{Hamiltonian}
The two QSHI rings $A$ and $B$ have their spin edge states described by single particle hamiltonians $\hat{H}_{A}$ and $\hat{H}_{B}$ ($\hbar=1$), respectively, given explicitly by 
%We consider two QSH insulator rings, the corresponding single particle hamiltonians
\begin{eqnarray}
\hat{H}_{A}&=&-iv_{F,A}\partial_{x_{j_{A}}}\sigma_{z} -i\alpha_{A}\partial_{x_{j_{A}}}\sigma_{y}-eV_{gA}, \label{ha} \\
\hat{H}_{B}&=&-iv_{F,B}\partial_{x_{j_{B}}}\sigma_{z} -i\alpha_{B}\partial_{x_{j_{B}}}\sigma_{y}-eV_{gB}, \label{hb}
\end{eqnarray}
where $v_{F,A}$ and $v_{F,B}$ are the Fermi velocities, $\alpha_{A}$ and $\alpha_{B}$ are the Rashba couplings, the variables 
$j_{A},j_{B}=1,2,...,7$ and coordinates $x_{1A},x_{2A},...,x_{7A},x_{1B},x_{2B},...,x_{7B}$ have associated 
different edge lenghts $l_{1A},l_{2A},...,l_{7A},l_{1B},l_{2B},...,l_{7B}$, $V_{gA}$ and $V_{gB}$ are the effective potentials associated to the each 
QSHI rings $A$ and $B$, respectively. Unitary operations 
\begin{eqnarray}
\hat{U}_{A}&=&\exp{\left(-i\left[\frac{\sigma_{x,A}}{2}\arcsin\left(\frac{\alpha_{A}}{v_{F,A}}
\right)\right]\right)} \\
\hat{U}_{B}&=&\exp{\left(-i\left[\frac{\sigma_{x,B}}{2}\arcsin\left(\frac{\alpha_{B}}{v_{F,B}}
\right)\right]\right)}
\end{eqnarray}
%$$ and $$ 
can be applied to the hamiltonians, eqs (\ref{ha}) and (\ref{hb}), in order to obtain
\begin{eqnarray}
\hat{H}_{A}&=&-iv_{\alpha,A}\partial_{x_{j_{A}}}\widetilde{\sigma}_{z,B} -eV_{gA}, \\
\hat{H}_{B}&=&-iv_{\alpha,B}\partial_{x_{j_{B}}}\widetilde{\sigma}_{z,B} -eV_{gB},
\end{eqnarray}
where now the Rashba terms are absorbed in the modified 
Fermi velocities 
\begin{eqnarray}
v_{\alpha,A}&=&\sqrt{v_{F,A}^{2}+\alpha_{A}^{2}},\\
v_{\alpha,B}&=&\sqrt{v_{F,B}^{2}+\alpha_{B}^{2}},
\end{eqnarray}
depending on the Rashba coupling in each QSHI ring, $\alpha_{A}$ and $\alpha_{B}$. The edge state momenta %are then given by
%the edge state momenta 
for the QSHI rings $A$ and $B$ are then given respectivelly by
\begin{eqnarray}
K_{A}&=& v_{\alpha, A}^{-1}\left(E_{A} + eV_{gA}\right), \\
K_{B}&=&v_{\alpha, B}^{-1}\left(E_{B} + eV_{gB}\right),
\end{eqnarray}
where $E_{A}$ and $E_{B}$ are the respective kinetic energies of each Krammers' partner in each QSHI ring.
%$$ 
%and $$
\section{Scattering matrices}
The relations between ingoing states, i.e., the states injected by the source, and the outgoing states, that one postselected by the detectors in the QSHI rings, are given by the standard procedure of Büttiker-Landau formalism \cite{buttiker1988}
\begin{eqnarray}
\hat{b}_{\sigma,j,A}&=&\sum_{\sigma'=\uparrow,\downarrow,j'=0,1,2}({\bf S})_{A,j\sigma\sigma'}\hat{a}_{\sigma'j',A},\\
\hat{b}_{\sigma,j,B}&=&\sum_{\sigma'=\uparrow,\downarrow,j'=0,1,2}({\bf S})_{B,j\sigma\sigma'}\hat{a}_{\sigma'j',B},
\end{eqnarray}
where in the index notation $j,j'$, the source is identified by the number $0$ %corresponds to the source $S$, 
the detector $D1$ by the number $1$ and the detector detector $D2$ by the number $2$, 
$\sigma, \sigma'$ corresponds to the spin states $\uparrow,\downarrow$ and the labels $A$ and $B$ correspond to in each QSHI ring. The scattering matrices $({\bf S})_{A,j\sigma\sigma'}$ and $({\bf S})_{B,j\sigma\sigma'}$ can be written in terms of the QSHI ring parameters by the following respective scattering matrices  
\begin{eqnarray}
\widetilde{({\bf S})}_{A}=\left( 
\begin{array}{cc}
p_{aA}e^{-i\varphi_{12A}} & f_{bA}t_{aA}^{*}e^{-i\varphi_{246A}} \\ 
f_{aA}^{*}t_{bA}e^{-i\varphi_{136A}} & p_{bA}^{*}e^{-i\varphi_{34A}}\\
-t_{aA}e^{-i\varphi_{15A}} & f_{bA}p_{aA}^{*}e^{-i\varphi_{456A}} \\ 
-f_{aA}^{*}p_{bA}e^{-i\varphi_{167A}} & t_{bA}^{*}e^{-i\varphi_{47A}}\\
-f_{aA}^{*}f_{bA}e^{-i\varphi_{146A}} & 0 \\ 
0 & -f_{aA}^{*}f_{bA}e^{-i\varphi_{146A}}
\end{array}%
\right) \label{smra}
\end{eqnarray}
and
\begin{eqnarray}
\widetilde{({\bf S})}_{B}=\left( 
\begin{array}{cc}
p_{aB}e^{-i\varphi_{12B}} & f_{bB}t_{aB}^{*}e^{-i\varphi_{246B}} \\ 
f_{aB}^{*}t_{bB}e^{-i\varphi_{136B}} & p_{bB}^{*}e^{-i\varphi_{34B}}\\
-t_{aB}e^{-i\varphi_{15B}} & f_{bB}p_{aB}^{*}e^{-i\varphi_{456B}} \\ 
-f_{aB}^{*}p_{bB}e^{-i\varphi_{167B}} & t_{bB}^{*}e^{-i\varphi_{47B}}\\
-f_{aB}^{*}f_{bB}e^{-i\varphi_{146B}} & 0 \\ 
0 & -f_{aB}^{*}f_{bB}e^{-i\varphi_{146B}}
\end{array}%
\right) \label{smrb}
\end{eqnarray}
where the phase parameters are written in terms of the lenghts of in each path of the QSHI rings. For the edge state momenta 
%\begin{eqnarray}
%K_{A}=v_{\alpha, A}^{-1}\left(E_{A} + eV_{gA}\right)
%\end{eqnarray}
$K_{A}$,  the phases in each length of the QSHI rings $A$ are given by
\begin{eqnarray}
\varphi_{12A}&=&K_{A}\left(l_{1A}+l_{2A}\right), \label{fi1}\\
\varphi_{15A}&=&K_{A}\left(l_{1A}+l_{5A}\right), \label{fi2}\\
\varphi_{34A}&=&K_{A}\left(l_{3A}+l_{4A}\right), \label{fi3}\\
\varphi_{47A}&=&K_{A}\left(l_{4A}+l_{7A}\right), \label{fi4}\\
\varphi_{136A}&=&K_{A}\left(l_{1A}+l_{3A}+l_{6A}\right), \label{fi5}\\
\varphi_{146A}&=&K_{A}\left(l_{1A}+l_{4A}+l_{6A}\right), \label{fi6}\\
\varphi_{167A}&=&K_{A}\left(l_{1A}+l_{6A}+l_{7A}\right), \label{fi7}\\
\varphi_{246A}&=&K_{A}\left(l_{2A}+l_{4A}+l_{6A}\right), \label{fi8}\\
\varphi_{456A}&=&K_{A}\left(l_{4A}+l_{5A}+l_{6A}\right). \label{fi9}
\end{eqnarray}
Similarly, for the edge state momenta 
%\begin{eqnarray}
%K_{B}=v_{\alpha, B}^{-1}\left(E_{B} + eV_{gB}\right),
%\end{eqnarray}
$K_{B}$, in the QSHI rings $B$, the phases in each length of the QSHI rings $B$ are given by
\begin{eqnarray}
\varphi_{12B}&=&K_{B}\left(l_{1B}+l_{2B}\right),\\
\varphi_{15B}&=&K_{B}\left(l_{1B}+l_{5B}\right),\\
\varphi_{34B}&=&K_{B}\left(l_{3B}+l_{4B}\right),\\
\varphi_{47B}&=&K_{B}\left(l_{4B}+l_{7B}\right),\\
\varphi_{136B}&=&K_{B}\left(l_{1B}+l_{3B}+l_{6B}\right),\\
\varphi_{146B}&=&K_{B}\left(l_{1B}+l_{4B}+l_{6B}\right),\\
\varphi_{167B}&=&K_{B}\left(l_{1B}+l_{6B}+l_{7B}\right),\\
\varphi_{246B}&=&K_{B}\left(l_{2B}+l_{4B}+l_{6B}\right),\\
\varphi_{456B}&=&K_{B}\left(l_{4B}+l_{5B}+l_{6B}\right).
\end{eqnarray}
In this simplified form, we can write the scattering processes by means of the operators
\begin{eqnarray}
\hat{b}_{\sigma,j,A}&=&\sum_{\sigma'=\uparrow,\downarrow}\widetilde{({\bf S})}_{A,j\sigma\sigma'}\hat{a}_{\sigma'0,A},\\
\hat{b}_{\sigma,j,B}&=&\sum_{\sigma'=\uparrow,\downarrow}\widetilde{({\bf S})}_{B,j\sigma\sigma'}\hat{a}_{\sigma'0,B},
\end{eqnarray}
and the generation of two particle states are then a result of the following operator relations
%The two particle states are 
\begin{eqnarray}
\hat{b}_{\sigma,j,A}\hat{b}_{\sigma''',j',A}&=&\sum_{\sigma'',\sigma'=\uparrow,\downarrow}\widetilde{({\bf S})}_{A,j\sigma\sigma'}\hat{a}_{\sigma'0,A}\widetilde{({\bf S})}_{A,j'\sigma'''\sigma''}\hat{a}_{\sigma''0,A}, \nonumber \\
&&  \label{bb1}\\
\hat{b}_{\sigma,j,A}\hat{b}_{\sigma''',j',A}&=&\sum_{\sigma'',\sigma'=\uparrow,\downarrow}\widetilde{({\bf S})}_{A,j\sigma\sigma'}\hat{a}_{\sigma'0,A}\widetilde{({\bf S})}_{A,j'\sigma'''\sigma''}\hat{a}_{\sigma''0,A}.\nonumber \\
&& \label{bb2}
\end{eqnarray}

\section{Quantum states in the QSHI rings %$A$ and $B$
}

The operator relations in the equations (\ref{bb1}) and (\ref{bb2}) imply that the two-particle states generated in the two QSHI rings $A$ and $B$, respectively, can be represented by
%
%In the case of a two particles these
\begin{eqnarray}
|\psi_{A}\rangle &=& \sum_{\sigma,\sigma'=\uparrow,\downarrow}A_{\sigma\sigma'}|\sigma\rangle_{1,A}|\sigma'\rangle_{2,A}, \label{qsra} \\
|\psi_{B}\rangle &=& \sum_{\sigma,\sigma'=\uparrow,\downarrow}B_{\sigma\sigma'}|\sigma\rangle_{1,B}|\sigma'\rangle_{2,B}, \label{qsrb}
\end{eqnarray}
where the amplitudes are obtained explicitly by using the reduced scattering matrix in eq. (\ref{smra}) for the QSHI ring A %and given by 
\begin{eqnarray}
A_{\uparrow\uparrow}&=& N_{A}f_{bA}\left(|p_{aA}|^{2}+|t_{aA}|^{2}\right)e^{-iK_{A}\left(l_{\uparrow\uparrow A}+l_{6A}\right)},\\
A_{\downarrow\downarrow}&=& N_{A}f_{aA}^{*}\left(|p_{bA}|^{2}+|t_{bA}|^{2}\right)e^{-iK_{A}\left(l_{\downarrow\downarrow A}+l_{6A}\right)},\\
A_{\downarrow\uparrow} &=& N_{A}[p_{bA}^{*}t_{aA}e^{-iK_{A}l_{\downarrow\uparrow A}}
\nonumber \\
&+& f_{aA}^{*}f_{bA}p_{aA}^{*}t_{bA}e^{-iK_{A}\left(l_{\downarrow\uparrow A}+2l_{6A}\right)}],\\
%&& \\
A_{\uparrow\downarrow} &=& N_{A}[p_{aA}t_{bA}^{*}e^{-iK_{A}l_{\uparrow\downarrow A}} \nonumber \\
&+& f_{aA}^{*}f_{bA}p_{bA}t_{aA}^{*}e^{-iK_{A}\left(l_{\uparrow\downarrow A}+2l_{6A}\right)}],%\nonumber \\
\end{eqnarray}
while for the QSHI ring B, using the reduced scattering matrix in eq. (\ref{smrb}), the amplitudes are given explicitly by 
\begin{eqnarray}
B_{\uparrow\uparrow}&=& N_{B}f_{bB}\left(|p_{aB}|^{2}+|t_{aB}|^{2}\right)e^{-iK_{B}\left(l_{\uparrow\uparrow B}+l_{6B}\right)},\\
B_{\downarrow\downarrow}&=& N_{B}f_{aB}^{*}\left(|p_{bB}|^{2}+|t_{bB}|^{2}\right)e^{-iK_{B}\left(l_{\downarrow\downarrow B}+l_{6B}\right)},\\
B_{\downarrow\uparrow} &=& N_{B}[p_{bB}^{*}t_{aB}e^{-iK_{B}l_{\downarrow\uparrow B}}
\nonumber \\
&+& f_{aB}^{*}f_{bB}p_{aB}^{*}t_{bB}e^{-iK_{B}\left(l_{\downarrow\uparrow B}+2l_{6B}\right)}],\\
%&& \\
B_{\uparrow\downarrow} &=& N_{B}[p_{aB}t_{bB}^{*}e^{-iK_{B}l_{\uparrow\downarrow B}}
\nonumber \\
&+& f_{aB}^{*}f_{bB}p_{bB}t_{aB}^{*}e^{-iK_{B}\left(l_{\uparrow\downarrow B}+2l_{6B}\right)}],%\nonumber \\
\end{eqnarray}
where the folloing effective path lengths are defined 
\begin{eqnarray}
l_{\uparrow\uparrow A}&=& l_{1A} + l_{2A} + l_{4A} + l_{5A}, \label{uu1} \\
l_{\uparrow\downarrow A}&=& l_{1A} + l_{2A} + l_{4A} + l_{7A}, \label{uu2} \\
l_{\downarrow\uparrow A}&=& l_{1A} + l_{3A} + l_{4A} + l_{5A}, \label{uu3}\\
l_{\downarrow\downarrow A}&=& l_{1A} + l_{3A} + l_{4A} + l_{7A}. \label{uu4} \\
l_{\uparrow\uparrow B}&=& l_{1B} + l_{2B} + l_{4B} + l_{5B}, \label{ub1} \\
l_{\uparrow\downarrow B}&=& l_{1B} + l_{2B} + l_{4B} + l_{7B}, \label{ub2} \\
l_{\downarrow\uparrow B}&=& l_{1B} + l_{3B} + l_{4B} + l_{5B}, \label{ub3}\\
l_{\downarrow\downarrow B}&=& l_{1B} + l_{3B} + l_{4B} + l_{7B}. \label{ub4} 
\end{eqnarray}
The following identities can be verified involving relations (\ref{uu1}), (\ref{uu2}), (\ref{uu3}) and (\ref{uu4}) and the phases present in the reduced density matrices (\ref{fi1}), ..., (\ref{fi9}) for the QSHI ring $A$
\begin{eqnarray}
K_{A}l_{\uparrow\downarrow A} &=& \varphi_{12A} + \varphi_{47A}, \\
K_{A}l_{\downarrow\uparrow A} &=& \varphi_{15A} + \varphi_{34A}, \\
K_{A}l_{\uparrow\uparrow A} + K_{A}l_{6A}&=& \varphi_{12A} + \varphi_{456A}, \\
K_{A}l_{\uparrow\downarrow A} + 2 K_{A}l_{6A} &=& \varphi_{167A} + \varphi_{246A}, \\
K_{A}l_{\downarrow\uparrow A} + 2 K_{A}l_{6A} &=& \varphi_{136A} + \varphi_{456A}, \\
K_{A}l_{\downarrow\downarrow A}+ K_{A}l_{6A} &=& \varphi_{34A} + \varphi_{167A}. %, \\
\end{eqnarray}
and similarly for the QSHI ring $B$ 
\begin{eqnarray}
K_{B}l_{\uparrow\downarrow B} &=& \varphi_{12B} + \varphi_{47B}, \\
K_{B}l_{\downarrow\uparrow B} &=& \varphi_{15B} + \varphi_{34B}, \\
K_{B}l_{\uparrow\uparrow B} + K_{B}l_{6B}&=& \varphi_{12B} + \varphi_{456B}, \\
K_{B}l_{\uparrow\downarrow B} + 2 K_{A}l_{6B} &=& \varphi_{167B} + \varphi_{246B}, \\
K_{B}l_{\downarrow\uparrow B} + 2 K_{A}l_{6B} &=& \varphi_{136B} + \varphi_{456B}, \\
K_{B}l_{\downarrow\downarrow B}+ K_{A}l_{6B} &=& \varphi_{34B} + \varphi_{167B}. %, \\
\end{eqnarray}
This allows to rewrite explicitly the amplitudes in terms of the phases present in the reduced scattering matrices in the QSHI ring $A$, given by
\begin{eqnarray}
A_{\uparrow\uparrow}&=& N_{A}f_{bA}\left(|p_{aA}|^{2}+|t_{aA}|^{2}\right)e^{-i\left(\varphi_{12A} + \varphi_{456A}\right)}, \label{auu}\\
A_{\downarrow\downarrow}&=& N_{A}f_{aA}^{*}\left(|p_{bA}|^{2}+|t_{bA}|^{2}\right)e^{-i\left( \varphi_{34A} + \varphi_{167A}\right)}, \label{add}\\
A_{\downarrow\uparrow} &=& N_{A}[p_{bA}^{*}t_{aA}e^{-i\left(\varphi_{15A} + \varphi_{34A}\right)} \label{adu}
\nonumber \\
&+& f_{aA}^{*}f_{bA}p_{aA}^{*}t_{bA}e^{-i\left(\varphi_{136A} + \varphi_{456A}\right)}],\\
%&& \\
A_{\uparrow\downarrow} &=& N_{A}[p_{aA}t_{bA}^{*}e^{-i\left(\varphi_{12A} + \varphi_{47A}\right)}
\nonumber \\
&+& f_{aA}^{*}f_{bA}p_{bA}t_{aA}^{*}e^{-i\left(\varphi_{167A} + \varphi_{246A}\right)}],\label{aud}
\end{eqnarray}
and for the QSHI ring $B$, given by
\begin{eqnarray}
B_{\uparrow\uparrow}&=& N_{B}f_{bB}\left(|p_{aB}|^{2}+|t_{aB}|^{2}\right)e^{-i\left(\varphi_{12B} + \varphi_{456B}\right)}, \label{buu}\\
B_{\downarrow\downarrow}&=& N_{B}f_{aB}^{*}\left(|p_{bB}|^{2}+|t_{bB}|^{2}\right)e^{-i\left( \varphi_{34B} + \varphi_{167B}\right)}, \label{bdd}\\
B_{\downarrow\uparrow} &=& N_{B}[p_{bB}^{*}t_{aB}e^{-i\left(\varphi_{15B} + \varphi_{34B}\right)}
\nonumber \\
&+& f_{aB}^{*}f_{bB}p_{aB}^{*}t_{bB}e^{-i\left(\varphi_{136B} + \varphi_{456B}\right)}], \label{bdu}\\
%&& \\
B_{\uparrow\downarrow} &=& N_{B}[p_{aB}t_{bB}^{*}e^{-i\left(\varphi_{12B} + \varphi_{47B}\right)}
\nonumber \\
&+& f_{aB}^{*}f_{bB}p_{bB}t_{aB}^{*}e^{-i\left(\varphi_{167B} + \varphi_{246B}\right)}],\label{bud} %\nonumber \\
\end{eqnarray}
%We can also rewrite these amplitudes using the relations
%\begin{eqnarray}
%A_{\uparrow\uparrow}&=& N_{A}f_{bA}\left(1-|f_{aA}|^{2}\right)e^{-i\left(\varphi_{12A} + \varphi_{456A}\right)},\\
%A_{\downarrow\downarrow}&=& N_{A}f_{aA}^{*}\left(1-|f_{bB}|^{2}\right)e^{-i\left( \varphi_{34A} + \varphi_{167A}\right)},\\
%A_{\downarrow\uparrow} &=& N_{A}[p_{bA}^{*}t_{aA}e^{-i\left(\varphi_{15A} + \varphi_{34A}\right)}
%\nonumber \\
%&+& f_{aA}^{*}f_{bA}p_{aA}^{*}t_{bA}e^{-i\left(\varphi_{136A} + \varphi_{456A}\right)}],\\
%&& \\
%A_{\uparrow\downarrow} &=& N_{A}[p_{aA}t_{bA}^{*}e^{-i\left(\varphi_{12A} + \varphi_{47A}\right)}
%\nonumber \\
%&+& f_{aA}^{*}f_{bA}p_{bA}t_{aA}^{*}e^{-i\left(\varphi_{167A} + \varphi_{246A}\right)}],%\nonumber \\
%\end{eqnarray}
%and for the QSHI ring $B$ given by
%\begin{eqnarray}
%B_{\uparrow\uparrow}&=& N_{B}f_{bB}\left(1-|f_{aB}|^{2}\right)e^{-i\left(\varphi_{12B} + \varphi_{456B}\right)},\\
%B_{\downarrow\downarrow}&=& N_{B}f_{aB}^{*}\left(1-|f_{bB}|^{2}\right)e^{-i\left( \varphi_{34B} + \varphi_{167B}\right)},\\
%B_{\downarrow\uparrow} &=& N_{B}[p_{bB}^{*}t_{aB}e^{-i\left(\varphi_{15B} + \varphi_{34B}\right)}
%\nonumber \\
%&+& f_{aB}^{*}f_{bB}p_{aB}^{*}t_{bB}e^{-i\left(\varphi_{136B} + \varphi_{456B}\right)}],\\
%&& \\
%B_{\uparrow\downarrow} &=& N_{B}[p_{aB}t_{bB}^{*}e^{-i\left(\varphi_{12B} + \varphi_{47B}\right)}
%\nonumber \\
%&+& f_{aB}^{*}f_{bB}p_{bB}t_{aB}^{*}e^{-i\left(\varphi_{167B} + \varphi_{246B}\right)}].%\nonumber \\
%\end{eqnarray} 
where $N_{A}$ and $N_{B}$ are the normalization factors satisfying %$\langle \psi_{A}|\psi_{A}\rangle = 1$ and $\langle \psi_{B}|\psi_{B}\rangle = 1$.
\begin{eqnarray}
&& \langle \psi_{A}|\psi_{A}\rangle = 1, \\
&& \langle \psi_{B}|\psi_{B}\rangle = 1,
\end{eqnarray}
provided by the corresponding normalization conditions.
%These normalization imply that the normalization facotors can be written as
%\begin{eqnarray}
%N_{A}&=& \left[|f_{bA}|^{2}\left(|t_{aA}|^{2}+|p_{aA}|^{2}\right)^{2} + |p_{aA}t^{*}_{bA} 
%+ f_{aA}^{*}f_{bA}p_{bA}t^{*}_{aA}e^{-}
%\end{eqnarray} 

\section{Quantum teleportation protocol}

Now the two QSHI rings $A$ and $B$ are considered in the states given respectively by equations (\ref{qsra}) and (\ref{qsrb}). %in the following form achieved by the control of the parameters in each QSHI ring [ref]
As was discussed in Sec I and II, the QSHI ring $A$ will be used as a qubit state, beloning to Alice, while the QSHI ring $B$ will be used as a quantum channel, shared between Alice and Bob. As a consequence, in the quantum teleportation protocol, the final state of Bob's state is in the QSHI ring $B$ and should be a teleported qubit from Alice's side, that consequently implements a quantum teleportation from the QSHI ring $A$ to QSHI ring $B$. The classical channel, also called feed-forward \cite{ma2012,steffen2013}, of the protocol is represented by communication of the outcome to Bob, allowing him to implement the appropriate adjustment of the tunneling junction in order to make possible, in the present scheme, the appropriate rotation of his state, ending up with a quantum teleported qubit.

In the next subsection, each subpart of the protocol will be detailed, what will make clear how it works in the scenario of two QSHI rings device. 

\subsection{Qubit generation}

By realizing a measument in the detector $D_{2A}$, with a state selection into the spin state $|\uparrow\rangle_{2A}$, the QSHI ring $A$ will have its original state filtered in the following qubit state of an edge state superposition in the mode ${1,A}$
\begin{eqnarray}
|\mathcal{Q}\rangle_{1,A}=A_{\uparrow\uparrow}|\uparrow\rangle_{1,A}+A_{\downarrow\uparrow}|\downarrow\rangle_{1,A}. \label{qbes}
\end{eqnarray}
This is a prototypical qubit state ready for be used in quantum information and quantum computation protocols \cite{nilsen}. The advantage of this qubit is that, since it is an edge state superposition, it is robust against backscattering pertubations \cite{hou2009}, maintaining the qubit useful against decoherence for longer times than other more usual systems \cite{zurek2003,groh2016}. This property also makes the edge states in QSHI rings very interesting for spintronics applications integrated with quantum information, in contrast to other systems \cite{lombardi2009}.

The coefficients $A_{\uparrow\uparrow}$ and $A_{\downarrow\uparrow}$ in eq. (\ref{qbes}) are explicitly given by equations (\ref{auu}) and (\ref{adu}).
%\begin{widetext}
%\begin{eqnarray}
%A_{\uparrow\uparrow}&=& N_{A}f_{Ab}\left(|p_{aA}|^{2}+|t_{aA}|^{2}\right)e^{-iK_{A}\left(l_{\uparrow\uparrow A}+l_{6A}\right)},\\
%A_{\downarrow\uparrow} &=& N_{A}\left[p_{bA}^{*}t_{aA}e^{-iK_{A}l_{\downarrow\uparrow A}}
%+f_{aA}^{*}f_{bA}p_{aA}^{*}t_{bA}e^{-iK_{A}\left(l_{\downarrow\uparrow A}+2l_{6A}\right)}\right].\nonumber \\
%\end{eqnarray} 
The control the tunneling junctions in the QSHI $A$ will project the qubit as a single quantum gate. In the present case however, these parameters are set free, although robust in the quantum superposition, in order to realize the quantum protocol in view. An important aspect to highlight is that, as a single qubit generator with low dissipation, the QSHI can also be used as a quantum memory for the qubit generated. 
%\end{widetext}

\subsection{Quantum channel}

After the generation of a qubit in the QSHI ring $A$,
%this last procedure, 
the system of the two QSHI rings $A$ and $B$ is described by the following total state
%\begin{widetext}
\begin{eqnarray}
|\mathcal{R}\rangle_{AB}&=&|\mathcal{Q}\rangle_{1,A}\otimes\left(\sum_{\sigma,\sigma'=\uparrow,\downarrow}B_{\sigma\sigma'}|\sigma\rangle_{1,B}|\sigma'\rangle_{2,B}\right), \label{totsab}
\end{eqnarray} 
%\end{widetext}
where now the QSHI ring $B$ has the role of a quantum channel in which the mode ${1,B}$ corresponds to an Alice's part and the mode ${2,B}$ corresponds to Bob's part of the quantum channel. As a consequence of the quantum teleportation protocol \cite{bennett1993}, Alice's qubit ${1,A}$ should be teleported to the Bob's mode ${2,B}$ as a qubit carrying the same coefficients of Alice's qubit. The result is the quantum teleportation process, Alice's qubit is destroyed in the mode ${1,A}$ and recreated in the mode ${2,B}$.  

The coefficients of the QSHI ring $B$, $B_{\uparrow\uparrow}$, $B_{\downarrow\downarrow}$, $B_{\downarrow\uparrow}$ and $B_{\uparrow\downarrow}$ in the equation (\ref{totsab}) are given explicitly by equations (\ref{buu}), (\ref{bdd}), (\ref{bdu}) and (\ref{bud}).
%\begin{eqnarray}
%B_{\uparrow\uparrow}&=& N_{B}f_{Bb}\left(|p_{Ba}|^{2}+|t_{Ba}|^{2}\right)e^{-iK_{B}\left(l_{B\uparrow\uparrow}+l_{B6}\right)},\\
%B_{\downarrow\downarrow}&=& N_{B}f_{Ba}^{*}\left(|p_{Bb}|^{2}+|t_{Bb}|^{2}\right)e^{-iK_{B}\left(l_{B\downarrow\downarrow}+l_{B6}\right)},\\
%B_{\downarrow\uparrow} &=& N_{B}\left[p_{Bb}^{*}t_{a}e^{-iK_{B}l_{B\downarrow\uparrow}}
%+f_{Ba}^{*}f_{Bb}p_{Ba}^{*}t_{Bb}e^{-iK_{B}\left(l_{B\downarrow\uparrow}+2l_{B6}\right)}\right],\nonumber\\
%&& \\
%B_{\uparrow\downarrow} &=& N_{B}\left[p_{Ba}t_{Bb}^{*}e^{-iK_{B}l_{B\uparrow\downarrow}}
%+f_{Ba}^{*}f_{Bb}p_{Bb}t_{Ba}^{*}e^{-iK_{B}\left(l_{B\uparrow\downarrow}+2l_{B6}\right)}\right],\nonumber \\
%\end{eqnarray} 

These coefficients can be propely adjusted by the control of the tunneling junction amplitudes \cite{strom2015}, allowing the generation of a maximally entangled quantum channel in the QSHI ring $B$.

\subsection{Bell type joint measurement}

The correlation between the qubit in the two QSHI ring $A$ and the quantum channel in the QSHI ring $B$ is achieved by means of a joint measurement detection in both rings. This type of operation have been largely used in quantum information protocols \cite{schuck2006,houwelingen2006} % and in particular in solid state scenario [ref]. 
Here the joint measurement is considered taking into account a common detector attaching the two QSHI rings $A$ and $B$, called $D_{1AB}$, corresponding to the simultaneous joint measurements of the modes ${1A}$ and ${1B}$ in a coincidence basis.%\cite{walborn2003}. 
This is equivalent to perform a joint measurement of Bell type in both outer edges $1A$ and $1B$, implemented by the action of one of the following Bell-type joint operators in the total state
\begin{eqnarray}
\hat{\mathcal{O}}^{\Phi(11)}_{\pm, AB}=|\Phi_{\pm}\rangle^{(1)}_{AB}\langle\Phi_{\pm}|^{(1)}_{AB}, \label{jsm}
\end{eqnarray}
\begin{eqnarray}
\hat{\mathcal{O}}^{\Psi(11)}_{\pm, AB}=|\Psi_{\pm}\rangle^{(1)}_{AB}\langle\Psi_{\pm}|^{(1)}_{AB}, \label{jsm2}
\end{eqnarray}
with equal probability, where the states $|\Phi_{\pm}\rangle^{(1)}_{AB}$ and $|\Psi_{\pm}\rangle^{(1)}_{AB}$ are joint states coupling the two QSHI rings $A$ and $B$, written explicitly by means of a Bell basis defined as
\begin{eqnarray}
|\Phi\rangle^{(11)}_{\pm AB}=\frac{|\uparrow\rangle_{1,A}|\uparrow\rangle_{1,B}\pm |\downarrow\rangle_{1,A}|\downarrow\rangle_{1,B}}{\sqrt{2}},
\end{eqnarray}
\begin{eqnarray}
|\Psi\rangle^{(11)}_{\pm AB}=\frac{|\uparrow\rangle_{1,A}|\downarrow\rangle_{1,B}\pm |\downarrow\rangle_{1,A}|\uparrow\rangle_{1,B}}{\sqrt{2}}.
\end{eqnarray}
The Bell state measurement will project Bob's state, or equivalently the total state of the two QSHI rings $A$ and $B$, in one of the following states
\begin{eqnarray}
|\mathcal{F}^{\Phi}\rangle_{\pm,B}&=& A_{\uparrow\uparrow}\sum_{\sigma'=\uparrow,\downarrow}B_{\uparrow\sigma'}|\sigma'\rangle_{2,B} \nonumber \\
&\pm& A_{\downarrow\uparrow}\sum_{\sigma'=\uparrow,\downarrow}B_{\downarrow\sigma'}|\sigma'\rangle_{2,B}.  \label{fstlp1}
\end{eqnarray}
\begin{eqnarray}
|\mathcal{F}^{\Psi}\rangle_{\pm,B}&=& A_{\uparrow\uparrow}\sum_{\sigma'=\uparrow,\downarrow}B_{\downarrow\sigma'}|\sigma'\rangle_{2,B} \nonumber \\
&\pm& A_{\downarrow\uparrow}\sum_{\sigma'=\uparrow,\downarrow}B_{\uparrow\sigma'}|\sigma'\rangle_{2,B}.  \label{fstlp2}
\end{eqnarray}
These states can also be written as
\begin{eqnarray}
|\mathcal{F}^{\Phi}\rangle_{\pm,B}&=& \left(A_{\uparrow\uparrow}B_{\uparrow\uparrow} \pm A_{\downarrow\uparrow}B_{\downarrow\uparrow}\right)|\uparrow\rangle_{2,B} \nonumber \\
&+& \left(A_{\uparrow\uparrow}B_{\uparrow\downarrow} \pm A_{\downarrow\uparrow}B_{\downarrow\downarrow}\right)|\downarrow\rangle_{2,B}  \label{fstlp1}
\end{eqnarray}
\begin{eqnarray}
|\mathcal{F}^{\Psi}\rangle_{\pm,B}&=& \left(A_{\uparrow\uparrow}B_{\downarrow\uparrow} \pm A_{\downarrow\uparrow}B_{\uparrow\uparrow}\right)|\uparrow\rangle_{2,B} \nonumber \\
&+& \left(A_{\uparrow\uparrow}B_{\downarrow\downarrow} \pm A_{\downarrow\uparrow}B_{\uparrow\downarrow}\right)|\downarrow\rangle_{2,B}.  \label{fstlp2}
\end{eqnarray}
The outcome of Alice's measurement will indicate which one of the states in equations (\ref{fstlp1}) and (\ref{fstlp2}) is the final Bob's state. 

\subsection{Feed-forward channel}

In order to allow Bob to operate on his state, Alice then should use a classical channel in order to inform Bob what was her outcome, this is achieved by a two bit communication \cite{bennett1993}, also called feed-forward communication \cite{steffen2013}. After that, Bob is capable of implementing the operation equivalent to the corresponding qubit rotation $\mathcal{R}_{i}^{-1}$, that will conclude successfully the quantum teleportation of Alice's qubit. 
As it is known, the choice of the joint-measurement corresponds to Alice's choice of how to project jointly her particles, what will lead to one of four possible outcomes. This imply that after this procedure, Bob has to adjust exactly to the way the measurement was performed, otherwise the fidelity will be considerably decreased and the protocol can be damaged. This is the requirement of a classical channel to intermediate the protocol between Alice and Bob. 

In the present case, the corresponding rotation $\mathcal{R}_{i}^{-1}$ is realized when Bob adjusts electrically\cite{strom2015} the tunneling amplitudes in the junctions of the QSHI ring $B$ with respect to Alice's outcome. For the states (\ref{fstlp1}), the tunneling junctions in the QSHI ring $B$ can be adjusted by reducing abruptly the coefficients responsible for the preserved spins in the outer edges.%, in such a way that there is no preserved spins in the junctions. 
This condition is given by the equations
\begin{eqnarray}
p_{aB}&=&p_{aB}^{*}=0, \label{pb1} \\
p_{bB}&=&p_{bB}^{*}=0. \label{pb2}
\end{eqnarray}
Additionaly, the coefficients responsible for spin tunneling are adjusted to fix tunneling amplitudes at the same probability amplitudes. This condition is given by
\begin{eqnarray}
%f_{bB}&=& f_{aB}^{*}=f_{B}, \label{ft1}\\
|t_{aB}|^{2}&=& |t_{bB}|^{2},%=|t_{B}|^{2}. 
\label{ft2}
\end{eqnarray}
while the coefficients responsible for spin flipping are adjusted to fix the flipping at the same probability amplitudes by with a phase difference given by
\begin{eqnarray}
f_{bB}&=& \pm f_{aB}^{*}, %= \pm f_{B}
\label{ft1}%\\
%|t_{aB}|^{2}&=& |t_{bB}|^{2}=|t_{B}|^{2}. \label{ft2}
\end{eqnarray}
depending on if the state is $|\mathcal{F}^{\Phi}\rangle_{\pm,B}$, i. e., $f_{bB}= f_{aB}^{*}$ for $|\mathcal{F}^{\Phi}\rangle_{+,B}$ and $f_{bB}=-f_{aB}^{*}$ for $|\mathcal{F}^{\Phi}\rangle_{-,B}$. The geometric phases should satisfy the congruence equation
\begin{eqnarray}
\left(\varphi_{12B} + \varphi_{456B}\right)\equiv \left(\varphi_{34B} + \varphi_{167B}\right)\mod 2\pi, \label{cg1}
\end{eqnarray}
what can be done from the beginning, by fixing the edge lengths. It is immediately satisfied in the case QSHI rings have a circular shape.
%The above procedure is a standard part of the electric control of the quantum state in the QSHI ring B [ref]. %In this case, the coefficients in the QSHI ring $B$ will be given by
%\begin{eqnarray}
%B_{\uparrow\uparrow} &=&  N_{B}f_{B}|t_{B}|^{2}e^{-iK_{B}\left(l_{B\uparrow\uparrow}+l_{B6}\right)},\\
%B_{\downarrow\downarrow} &=& N_{B}f_{B}|t_{B}|^{2}e^{-iK_{B}\left(l_{B\uparrow\uparrow}+l_{B6}\right)}, \\
%B_{\downarrow\uparrow} &=& B_{\uparrow\downarrow}= 0.
%\end{eqnarray} 
Since now $B_{\downarrow\uparrow}= B_{\uparrow\downarrow}=0$ and $B_{\uparrow\uparrow}= \pm B_{\downarrow\downarrow}$ are satisfied, to the respective outcome $|\mathcal{F}^{\Phi}\rangle_{\pm,B}$, the term $B_{\uparrow\uparrow}$ can be simply removed from the quantum spin Hall ring $B$ by normalization, meaning that the QSHI ring B has been projected in the qubit state 
\begin{eqnarray}
|\mathcal{Q}\rangle_{2,B}=A_{\uparrow\uparrow}|\uparrow\rangle_{2,B} + A_{\downarrow\uparrow}|\downarrow\rangle_{2,B},
\end{eqnarray}
that has the exact form as Alice's original qubit. As a consequence, the Alice's qubit state in the mode ${1,A}$ was quantum teleported to Bob's side state in the mode ${2,B}$. In other words, the qubit originally in the QSHI ring $A$ is now teleported to the QSHI ring $B$. 

On the other hand, for the states (\ref{fstlp2}), the tunneling junctions in the QSHI ring $B$ can be adjusted by reducing abruptly the coefficients responsible for the flipped spins in the inner edges. %, in such a way that there is no preserved spins in the junctions. 
This condition is given by the relations
\begin{eqnarray}
f_{aB}&=& f_{aB}^{*}=0, \label{fb1} \\
f_{bB}&=& f_{bB}^{*}=0. \label{fb2}
\end{eqnarray}
Additionaly, the coefficients responsible for spin tunneling are adjusted to fix tunneling amplitudes under the condition
\begin{eqnarray}
%f_{bB}&=& f_{aB}^{*}=f_{B}, \label{ft1}\\
t_{aB}&=& t_{bB}^{*},%=|t_{B}|^{2}. 
\label{ft3}
\end{eqnarray}
what also leads to the condition of same probability amplitudes in equation (\ref{ft2}). This result means that Bob can fix this condition for any outcome.

The coefficients responsible for spin preservation at the same edge are adjusted to fix the preservation at the same probability amplitudes with a phase difference given by
\begin{eqnarray}
p_{bB}&=& \pm p_{aB}^{*}, %= \pm f_{B}
\label{ptb1}%\\
%|t_{aB}|^{2}&=& |t_{bB}|^{2}=|t_{B}|^{2}. \label{ft2}
\end{eqnarray}
depending on if the state is $|\mathcal{F}^{\Psi}\rangle_{\pm,B}$, i. e., $p_{bB}= p_{aB}^{*}$ for $|\mathcal{F}^{\Psi}\rangle_{+,B}$ and $p_{bB}=-p_{aB}^{*}$ for $|\mathcal{F}^{\Psi}\rangle_{-,B}$. In this case, The geometric phases should satisfy the congruence equation
\begin{eqnarray}
\left(\varphi_{15B} + \varphi_{34B}\right)\equiv \left(\varphi_{12B} + \varphi_{47B}\right)\mod 2\pi. \label{cg2}
\end{eqnarray}
This can be realized by fixing the lengths from the beginning, as in the congruence equation (\ref{cg1}). 

With these results, a precise procedure should be implemented by Bob after Alice's communication of her outcome. The table 1 summarizes the procedure that should be realized by Bob.
%\begin{table}[]
%\begin{tabular}{lllll}
% 1 & \| & 1 &  &  \\
% &  &  &  &  \\
% &  &  &  &  \\
% &  &  &  & 
%\end{tabular}
%\end{table}

\begin{center}
\textcolor{blue}{\bf Table 1: 
{\it Bob's adjust with respect to Alice's outcome}}
\scalebox{0.85}[1.0]{
\begin{tabular}{ |c|c|c|c| } 
 \hline
 {\bf Alice's outcome} & {\bf flipping} & {\bf tunneling} & {\bf preserving} \\ 
 \hline
 $|\Phi\rangle^{(11)}_{+ AB}$ & $f_{bB}= + f_{aB}^{*}$ & $t_{aB}=t_{bB}^{*}$ & $p_{aB}=p_{bB}=0$ \\ 
 \hline
$|\Phi\rangle^{(11)}_{- AB}$ & $f_{bB}= - f_{aB}^{*}$ & $t_{aB}=t_{bB}^{*}$ & $p_{aB}=p_{bB}=0$ \\ 
 \hline
$|\Psi\rangle^{(11)}_{+ AB}$ & $f_{bB}= f_{aB}=0$ & $t_{aB}=t_{bB}^{*}$ & $p_{bB}= + p_{aB}^{*}$ \\ 
 \hline
$|\Psi\rangle^{(11)}_{- AB}$ & $f_{bB}= f_{aB}=0$ & $t_{aB}=t_{bB}^{*}$ & $p_{bB}= - p_{aB}^{*}$ \\ 
 \hline
\end{tabular}
}
\end{center}

\subsection{Alternative protocol}

The choice of the measurement in the detector $D_{2A}$ is not arbitrary and then can be realized
alternatively with the measurement detector $D_{2A}$ projecting the mode ${2,A}$ into the state $|\downarrow\rangle_{2,A}$, case where the following qubit state is achieved
\begin{eqnarray}
|\tilde{\mathcal{Q}}\rangle_{1,A}=A_{\uparrow\downarrow}|\uparrow\rangle_{1,A} + A_{\downarrow\downarrow}|\downarrow\rangle_{1,A}. \label{qubit1a}
\end{eqnarray}
The coefficients in this case are $A_{\uparrow\downarrow}$ and $A_{\downarrow\downarrow}$, in contrast to the previous case, where the coefficients were $A_{\uparrow\uparrow}$ and $A_{\downarrow\uparrow}$, respectively. This choice reflects the fact the QSHI ring $A$ is producing a two-particle state, allowing two complementary ways of generating the qubit state. The next procedure is similar to the discussed before, where a joint measurement is realized and a classical communication is done, allowing to implement an equivalent procedure of quantum teleportation, using a distinct input qubit. %similar to the table 1.
%
%The coefficients in (\ref{qubit}) are given explictly in terms of the probability amplitudes [ref] by the following expression
%\begin{eqnarray}
%A_{\uparrow\downarrow} &=& N_{A}\left[p_{Aa}t_{Ab}^{*}e^{-iK_{A}l_{A\uparrow\downarrow}}
%+f_{Aa}^{*}f_{Ab}p_{Ab}t_{Aa}^{*}e^{-iK_{A}\left(l_{A\uparrow\downarrow}+2l_{A6}\right)}\right].\nonumber \\
%&&\\
%A_{\downarrow\downarrow}&=& N_{A}f_{Aa}^{*}\left(|p_{Ab}|^{2}+|t_{Ab}|^{2}\right)e^{-iK_{A}\left(l_{A\downarrow\downarrow}+l_{A6}\right)},
%\end{eqnarray}
In this case, the whole system is described by the following state of the QSHI rings $A$ and $B$
%\begin{widetext}
\begin{eqnarray}
|\tilde{\mathcal{R}}\rangle_{AB}&=&|\tilde{\mathcal{Q}}\rangle_{1,A}\otimes\left(\sum_{\sigma,\sigma'=\uparrow,\downarrow}B_{\sigma\sigma'}|\sigma\rangle_{1,B}|\sigma'\rangle_{2,B}\right), \nonumber \\
\end{eqnarray} 
where again the Alice's measurement of Bell type will project her states in one of the four Bell basis described before, whose outcome should be feed-forwarded to Bob, with a classical communication channel, and then allowing Bob to implement the quantum teleportation successfully. 
%\begin{eqnarray}
%|\widetilde{\mathcal{F}}\rangle_{B}&=&\sum_{\sigma'=\uparrow,\downarrow}A_{\uparrow\downarrow}B_{\uparrow\sigma'}|\sigma'\rangle_{2,B} 
%+ A_{\downarrow\downarrow}B_{\downarrow\sigma'}|\sigma'\rangle_{2,B}\nonumber \\
%\end{eqnarray}
Again, the joint measurement by Alice imply in how Bob's side should control the tunneling junctions in the QSHI ring $B$.
The Bell state measurement will project Bob's state, or equivalently the total state of the two QSHI rings $A$ and $B$, in one of the following states
\begin{eqnarray}
|\tilde{\mathcal{F}}^{\Phi}\rangle_{\pm,B}&=& A_{\uparrow\downarrow}\sum_{\sigma'=\uparrow,\downarrow}B_{\uparrow\sigma'}|\sigma'\rangle_{2,B} \nonumber \\
&\pm& A_{\downarrow\downarrow}\sum_{\sigma'=\uparrow,\downarrow}B_{\downarrow\sigma'}|\sigma'\rangle_{2,B}.  \label{fstlp1a}
\end{eqnarray}
\begin{eqnarray}
|\tilde{\mathcal{F}}^{\Psi}\rangle_{\pm,B}&=& A_{\uparrow\downarrow}\sum_{\sigma'=\uparrow,\downarrow}B_{\downarrow\sigma'}|\sigma'\rangle_{2,B} \nonumber \\
&\pm& A_{\downarrow\downarrow}\sum_{\sigma'=\uparrow,\downarrow}B_{\uparrow\sigma'}|\sigma'\rangle_{2,B}.  \label{fstlp2a}
\end{eqnarray}
These states can also be written as
\begin{eqnarray}
|\tilde{\mathcal{F}}^{\Phi}\rangle_{\pm,B}&=& \left(A_{\uparrow\downarrow}B_{\uparrow\uparrow} \pm A_{\downarrow\downarrow}B_{\downarrow\uparrow}\right)|\uparrow\rangle_{2,B} \nonumber \\
&+& \left(A_{\uparrow\downarrow}B_{\uparrow\downarrow} \pm A_{\downarrow\downarrow}B_{\downarrow\downarrow}\right)|\downarrow\rangle_{2,B}  \label{fstlp3a}
\end{eqnarray}
\begin{eqnarray}
|\tilde{\mathcal{F}}^{\Psi}\rangle_{\pm,B}&=& \left(A_{\uparrow\downarrow}B_{\downarrow\uparrow} \pm A_{\downarrow\downarrow}B_{\uparrow\uparrow}\right)|\uparrow\rangle_{2,B} \nonumber \\
&+& \left(A_{\uparrow\downarrow}B_{\downarrow\downarrow} \pm A_{\downarrow\downarrow}B_{\uparrow\downarrow}\right)|\downarrow\rangle_{2,B}.  \label{fstlp4a}
\end{eqnarray}
In this case, the quantum teleportation is achieved with the control of the spin preservation, flipping and tunneling amplitudes, where in this alternative situation, where the QSHI ring $B$ will be projected into the qubit state
\begin{eqnarray}
|\tilde{\mathcal{Q}}\rangle_{2,B}=A_{\uparrow\downarrow}|\uparrow\rangle_{2,B} + A_{\downarrow\downarrow}|\downarrow\rangle_{2,B},
\end{eqnarray}
that corresponds to the original amplitudes of the qubit in the QSHI ring $A$. 

This means again that Alice's qubit is also quantum teleported to Bob in the mode ${2,B}$. The quantum teleportation from the QSHI ring $A$ to the QSHI ring $B$ also achieved in this situation. 

Although the coefficients of the QSHI ring $A$ are changed, Bob should use the same table 1 to adjust his tunneling junctions in order to adjust his state. It follows that independently of the initial qubit, Bob can use the same table, i.e., table 1, to adjust the tunneling junctions in QSHI ring $B$ in order to obtain his particle. This result is important due the fact that the two QSHI rings can be realized with independent functions, one for qubit generation and the other to generate the quantum channel, with the use of the same table to implement the quantum teleportation.

\section{Bell-state analysis %Experimental perspectives
}

\begin{figure}[h]
\centering
\includegraphics[scale=0.25]{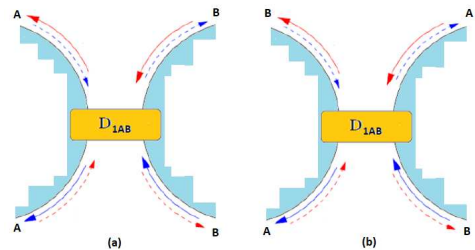}
\caption{(Color online) At the joint detector $D_{1AB}$ the edge states are identical, allowing two indistinguishable configurations 
(a) $A \rightarrow A$ and $B \rightarrow B$ or (b) $A \rightarrow B$ and $B \rightarrow A$.}
\label{2ringsbell}
\end{figure}

As realized in the first experimental proposal of quantum teleportation \cite{bouwmeester1997}, the main challenge in execute experimentally a quantum teleportation is the realization of the Bell state analysis. This is related to the measurement of a two-particle system \cite{weinfurter1994}. The projection into Bell states is achieved by making the two-particle states take indistinguishable configurations. This can be done in the case of two QSHI rings. In fact at the joint detector $D_{1AB}$ the edge states from the outer edge of the QSHI ring $A$ and from the outer edge of the QSHI ring $B$ will take two indistinguishable configurations in the intersection where the joint detector is present: The edge states can stay at the same QSHI ring
(a) $A \rightarrow A$ and $B \rightarrow B$ or they can move from one QSHI ring to the other (b) $A \rightarrow B$ and $B \rightarrow A$. This situation is displayed in figure \ref{2ringsbell}. This means that the set of edge states in the joint detector are indistinguishable as a set of identical particles.

%\subsection{Bell state analysis}

A joint measurement corresponds to measure the spin degrees of freedom in a rotated basis of the two states corresponding to the Bell basis, also called coincidence basis, corresponding to coincidences at different detectors. This measurement gives an incomplete information with respect to the computational basis description of the two-particle state 
\begin{eqnarray}
\lbrace |\uparrow\rangle_{1,A}|\uparrow\rangle_{1,B}, |\downarrow\rangle_{1,A}|\downarrow\rangle_{1,B}, |\uparrow\rangle_{1,A}|\downarrow\rangle_{1,B}, 
|\tilde{\downarrow}\rangle_{1,A}|\tilde{\uparrow}\rangle_{1,B}\rbrace, \nonumber
\end{eqnarray}
but gives a complete information with respect to the Bell basis 
\begin{eqnarray}
\lbrace |\tilde{\uparrow}\rangle_{1,A}|\tilde{\uparrow}\rangle_{1,B}, |\tilde{\downarrow}\rangle_{1,A}|\tilde{\downarrow}\rangle_{1,B}, |\tilde{\uparrow}\rangle_{1,A}|\tilde{\downarrow}\rangle_{1,B}, 
|\tilde{\downarrow}\rangle_{1,A}|\tilde{\uparrow}\rangle_{1,B} \rbrace. \nonumber
\end{eqnarray}
In this basis, the four Bell states are written simply as
\begin{eqnarray}
|\Phi\rangle^{(11)}_{+ AB}&=& |\tilde{\uparrow}\rangle_{1,A}|\tilde{\uparrow}\rangle_{1,B}, \\
|\Phi\rangle^{(11)}_{- AB}&=& |\tilde{\downarrow}\rangle_{1,A}|\tilde{\downarrow}\rangle_{1,B}, \\ 
|\Psi\rangle^{(11)}_{+ AB}&=& |\tilde{\uparrow}\rangle_{1,A}|\tilde{\downarrow}\rangle_{1,B},  \\
|\Psi\rangle^{(11)}_{- AB}&=& |\tilde{\downarrow}\rangle_{1,A}|\tilde{\uparrow}\rangle_{1,B},
\end{eqnarray}
that can be called a coincidence description of Bell states. It describes completely the two-particle Hilbert space for the edge states in the outer edges of QSHI rings $A$ and $B$. A tilde on top of the arrows of spins means that the state is not the spin state, but a coincidence state constituting a given Bell state. A Bell state measurement then consists in making a simultaneous measurement in the coincidence states $|\tilde{\uparrow}\rangle_{1,A}$ and $|\tilde{\uparrow}\rangle_{1,B}$, $|\tilde{\downarrow}\rangle_{1,A}$ and $|\tilde{\uparrow}\rangle_{1,B}$, $|\tilde{\uparrow}\rangle_{1,A}$ and $|\tilde{\downarrow}\rangle_{1,B}$, or 
$|\tilde{\downarrow}\rangle_{1,A}$ and $|\tilde{\downarrow}\rangle_{1,B}$, leading to four precise results. In the table 2, we display the frames with respect to each basis, where the normalization is omitted in the computational basis for simplicity. 
\begin{center}
\textcolor{blue}{\bf Table 2: {\it Frames with respect to Bell basis and Computational basis}}
\scalebox{0.85}[1.0]{
\begin{tabular}{ |c|c|c| } 
 \hline
 {\bf Bell state} & {\bf Coincidence frame} & {\bf Computational basis frame}  \\ 
 \hline
 $|\Phi\rangle^{(11)}_{+ AB}$ & $|\tilde{\uparrow}\rangle_{1,A}|\tilde{\uparrow}\rangle_{1,B}$ & $|\uparrow\rangle_{1,A}|\uparrow\rangle_{1,B} + |\downarrow\rangle_{1,A}|\downarrow\rangle_{1,B}$  \\ 
 \hline
$|\Phi\rangle^{(11)}_{- AB}$ & $|\tilde{\downarrow}\rangle_{1,A}|\tilde{\downarrow}\rangle_{1,B}$ & $|\uparrow\rangle_{1,A}|\uparrow\rangle_{1,B} - |\downarrow\rangle_{1,A}|\downarrow\rangle_{1,B}$ \\ 
 \hline
$|\Psi\rangle^{(11)}_{+ AB}$ & $|\tilde{\uparrow}\rangle_{1,A}|\tilde{\downarrow}\rangle_{1,B}$ & $|\uparrow\rangle_{1,A}|\downarrow\rangle_{1,B} + |\downarrow\rangle_{1,A}|\uparrow\rangle_{1,B}$   \\ 
 \hline
$|\Psi\rangle^{(11)}_{- AB}$ & $|\tilde{\downarrow}\rangle_{1,A}|\tilde{\uparrow}\rangle_{1,B}$ & $|\uparrow\rangle_{1,A}|\downarrow\rangle_{1,B} - |\downarrow\rangle_{1,A}|\uparrow\rangle_{1,B}$\\ 
 \hline
\end{tabular}
}
\end{center}
\begin{figure}[h]
\centering
\includegraphics[scale=0.3]{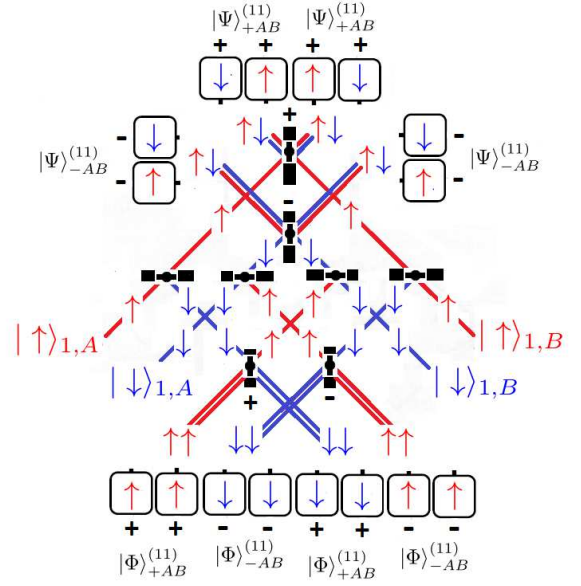}
\caption{(Color online) Bell state analysis at the joint detector $D_{1AB}$: The edge states are transmitted and reflected by solid-state beam splitters made of tunneling junctions or Zeeman-split quantum dots, leading to four distinguishable configurations, the four Bell states. The double lines corresponds to the Bell states, that are measured by coincidence detectors. Spin up is represented by red line, spin down is represented by blue line.}
\label{2ringsbell2}
\end{figure}
\begin{figure}[h]
\centering
\includegraphics[scale=0.25]{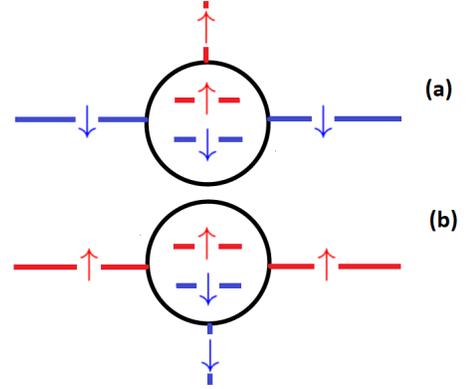}
\caption{(Color online) Solid-state beam-splitter (SSBP): %A tunneling junction defect or a Zeeman-split quantum dot. %The quantum dots are connected by edge states. Since the edge states do not backscatter, two leads conduct forwardly propagating and counter-propagating edge states states in different channels. %
In (a) a spin down interacts with the SSBP leading to a propagating (transmitted) or counter-propagating (reflected) spin. In (b) a spin up interacts with the SSBP leading to a propagating (transmitted) or counter-propagating (reflected) spin.}
\label{2ringqdots}
\end{figure}
A measurement in the Bell basis is achieved by means of a set of beam splitters followed by coincidence detectors \cite{weinfurter1994}. 

The figure \ref{2ringsbell2} displays the arrangement of implementation of the Bell state analysis, in which a set of beam splitters allows the detection of the four Bell states, using coincidence detectors. In the case of edge states these beam splitters can be tunneling junctions with fixed tunneling parameters, as the tunneling junctions connecting the inner and outer edges of the QSHI ring, or can be made of Zeemann-split quantum dots  \cite{probst2015}. 
For the case of a Zeemann-split quantum dot (ZSQD) 
the technique for the beam splitting of the edge states should be more improved, since it is necessary a %In the figure \ref{2ringqdots} is described 
a two-level ZSQD, where the opposite spins generates a more stable configuration, such that when a spin is injected in the ZSQD, there is the onset of a unstable regime leading to the emission of a single spin in up or down state, what can occur in both cases: (a) when the injection up and (b) when the injection is down, as it is displayed in the figure \ref{2ringqdots}. In a beam-splitter made of the ZSQDs, the outgoing edge states are forwardly moved to propagating or counterpropagating channels.%, since backscattering is forbidden in QSHIs. A balistic colision with the ZSQDs forbids the formation of excited state, leading to detection of the corresponding spin states in the corresponding coincidence basis, necessary to the Alice's joint measurement. The interaction between the edge states and the Zeemann-split quantum dots are then in dispersive regime [ref], with the quantum dots left in the ground state after the ballistic scattering [ref].
On the other hand, in the case of tunneling junctions, (a) when the injection up and (b) when the injection is down, the tunnel defect produces probability amplitudes of tunneling in propagating or counterpropagating states. The complete set of transmissions and reflections states will then generate a set of coincidence measurements, allowing the realization of Bell state analysis. The coincidence detectors are then coincidence detectors measuring the outgoing spin states.  
%A two level ZSQD in a stable situation of two levels with opposite spin, while the other situation imply the emission of a spin state in up or down state. By linking the ZSQD with QSHIs, allows that the spin emmission from the ZSQD should be injected in the 
%With the inset of the edge state in the ZSQD, th  
%
%both with the same confirguration of the figure .  and 

%The joint detector $D_{1AB}$ should implement a standard scheme for Bell state analysis applied to a scenario of this two QSHI rings device. 

%The two outer edge states comming from the QSHI rings $A$ and $B$ are splitted forwardly by transmited and reflected edge modes after balistic colisions with the ZSQDs in Coulomb blockade regime (or tunneling junctions with $50\%$  transmission and $50\%$ reflection) and are then jointly detected by coincidence measurement detectors that can exhibit one of the four Bell states as a result. 
The Bell state analysis allows Alice to distinguish each of the four results of the Bell state and then feedforward the result to Bob.%, allowing , as it is depicted in figure \ref{2ringsbell2}. 
Other possible implementations of a Bell state analysis for the the joint detector $D_{1AB}$ can be possible, although not considered in this proposal. Considering the total size of the two QSHI device described in the Sec II, the ZSQDs or the tunneling defects in the joint detector can have an estimated diameters with ranges 
from 2nm to 10nm, what could also achieved with use of nanocrystals \cite{barnejee2017}.

\section{Concluding remarks}

%In this paper, a 
The quantum teleportation between QSHI rings, %was proposed, where the discussions of 
qubit generation, entanglement generation and Bell state analysis in the context of edge states are discussed in this proposal. Important consequences of the results are for quantum information protocols in the domain of QSHI rings,  topological insulators, as well as spintronics. The two QSHI rings device discussed can be fabricated with an estimated total area of $230 \time 450$ nm$^{2}$, while the detectors involved can be built also in nanosize. These characteristics make this system ideal for quantum technologies integrated in other electronic and spintronic systems. Due its compact geometry and size parameters, the QSHI rings can be used to storage of qubits, generation of entangled states and transmission via quantum teleportation. This possibility makes such a device highly practical as a quantum device, with potential technological applications. It can also be verified by experimentally in the domain of QSHI physics.%, as other quantum teleportation protocol in other contexts [ref]. 
This proposal also sheds new light in the relation among quantum information protocols, topological insulators and spintronics, particularly in the domain of edge states. In the domain of solid state devices, it can be integrated with other solid state schemes involving different components used in modern technologies, allowing to integrate nanotechnologies, semiclassical devices, spintronic devices and quantum computers \cite{huo2018,quan2018,llewellyn2019}. It can then give a new routes for technologies emerging from nanostructured materials \cite{whitesides2005,zhao2017}, nanocrystals \cite{barnejee2017}, metamaterials \cite{rakhmanov2008,prudencio2018}%, %quantum machine learning \cite{canabarro2019} 
and teleportation networks \cite{huang2020}. 

\section{Acknowledgements}
The author is grateful for the valuable discussions with %A. Ström, in the early stages of this project, 
Prof. Henrik Johannesson. 
%that also made a critical reading of this manuscript. 
The author also would like to thank the Department of Physics at University of Gothenburg/Chalmers University of Technology, Gothenburg, Sweden, that provided the space and hospitality to make possible the conclusion of most part of this paper. %in the beautiful city of Gothenburg, Sweden. 
The partial support by CNPQ (Brazil) is also acknowledged.\\ %The author also thanks FAPEMA (Brazil), that supported the early stages of this project. 
%
%, and the indirect support by Swedish Academy of Sciences and Brazilian Academy of Sciences.%, H. Johannesson 
%is also acknowleged. %The author also thanks Anders Ström for valuable discussions in the early stages of this project.
%the previous support %by %FAPEMA (Brazil) APCInter project, 
%where this project has begun, 
%the partial support by CNPQ. The author thanks the University of Gothenburg for the nice stay during the visiting researcher 
\\
{\bf Data Availability}\\
The data used to support the findings of this study are
included within the article.\\
\\
{\bf Conflicts of Interest}\\
The author declares there is no conflicts of interest in this
paper.\\

\end{document}